\documentclass{article}
\usepackage{pdfpages}
\usepackage{xcolor}
\usepackage{graphicx}
\usepackage{cite}
\usepackage{caption}
\usepackage{subcaption}
\usepackage{authblk}
\usepackage{amsmath}
\usepackage[top=5cm, bottom=5cm, left=3cm, right=2cm, headsep=1.5cm, heightrounded=true]{geometry}
\usepackage{fancyhdr}
\graphicspath{ {{images/}} }

\title{  Black holes  thermodynamics with  CFT re-scaling    }
	\author[1,2]{Yahya Ladghami\thanks{ \texttt{yahya.ladghami@ump.ac.ma}}}
		\author[1,2]{Taoufik Ouali\thanks{ \texttt{t.ouali@ump.ac.ma}}}
	\affil[1] {Laboratory of Physics of Matter and Radiation, Mohammed I University, BP 717, Oujda, Morocco}
	\affil[2]{Astrophysical and Cosmological Center,  Faculty of Sciences, BP 717,  Oujda, Morocco}
\begin{document}
\maketitle
\begin{abstract}
In this paper, we study the thermodynamic behavior  of charged AdS black holes in a conformal holographic extended thermodynamic. Our setup is constructed using a new dictionary that relates AdS black hole quantities to the corresponding dual conformal fields theory (CFT) one, with the conformal factor being treated as a variable thermodynamic. In this thermodynamic study, we investigate the critical phenomena of charged AdS black holes and their relationship to the central charge value, C. Additionally, we examine the phase transitions and black holes stability using the free energy and the heat capacity, respectively. Furthermore, by examining the chemical potential, we  establish  a criteria that differentiate between quantum and classical black hole’s behaviors. Our setup highlights one of the key findings, namely traditional black hole thermodynamics acts as a boundary between  quantum and classical regimes.
\end{abstract}
\section{Introduction}
Black hole thermodynamics is a fundamental field of study in theoretical physics that has garnered significant attention. This interest arises from its ability to elucidate the connection between gravity and quantum physics  \cite{BH1}. The exploration of black hole thermodynamics commenced with Hawking and a group of physicists introducing a new perspective on our comprehension of black holes. They uncovered that black holes emit radiation, a phenomenon referred to as black holes evaporation in the literature, exhibit temperature and entropy \cite{i2,i3,i4}. Additionally, they formulated four laws analogous to those  of ordinary  thermodynamics \cite{i1}. But the process of a black hole evaporation results in the loss of information, a puzzling phenomenon that presents a profound challenge to the fundamental properties of quantum theory \cite{IN, IN1,IN2}. This conundrum raises questions about the conservation of information in the context of black holes.  Another important phenomenon related to black holes in anti-de Sitter (AdS) spacetime is the Hawking-Page phase transition \cite{HP}, which describes a transition between a black hole state and a pure thermal AdS. This transition can be interpreted as a confinement/deconfinement phase transition in the dual quark-gluon plasma \cite{BH2}. Additionally, thermodynamic of  black holes  with negative entropy \cite{SO1}, quantum corrections to thermodynamics of black holes \cite{SO2,SO3}, anti-evaporation of black holes in $F(R)$ gravity \cite{SO4,SO5}, shadow of the
black holes \cite{SO6} were also studied. In the literature, this kind of black holes thermodynamics is referred to as “the traditional black hole thermodynamics (TBHT)".
\\

In the last decade,  thermodynamics of black holes has witnessed significant development, due to the inclusion of the cosmological constant as a thermodynamic variable, interpreted as thermodynamic pressure and its conjugate quantity as a thermodynamic volume.  This approach is called “extended phase space thermodynamics (EPST)" or “black holes chemistry" \cite{Kastor,K1,K2,K3,K4,K5}. Within this framework, many phenomena and behaviors of black holes have been discovered, including critical phenomena,  phase transitions between small and large black holes, and their analogy with Van der Waals fluids \cite{PV,PV1,PV2,PV3,PV4,PV5,PV6}.  Additionally, polymer-type phase transitions \cite{T1}, stretched quintessence phase \cite{i12} multi-critical points \cite{T2,T3}, snapping transitions \cite{T4}, superfluid \cite{T5}, Joule-Thomson expansion of black holes \cite{T6,T7,T8,T9}, EGB black holes \cite{ma1,ma2,ma3,ma4,ma5,ma6,ma7,ma8} were also analyzed. Despite all this results, Gao and Zhao \cite{gd,s} recently pointed out some  problems that the EPST formalism suffers from, such as the problem of the “ensemble of theories", the non-homogeneity of  Smarr relation, and the interpretation of black holes mass as enthalpy. They then proposed another formalism competing with the EPST one, known as “restricted phase space thermodynamics (RPST).
\\

Building upon Maldacena duality or Anti-de Sitter/Conformal Field Theory (AdS/CFT) correspondence \cite{AdS}, Visser introduced a new thermodynamic framework that incorporates central charge, $C$, and chemical potential, $\mu$, as novel  thermodynamic variables \cite{vt}. However, it should be noted that Visser was not the first to include $C$ and $\mu$, as they  were introduced previously in \cite{vt1,vt2,vt3,vt4}. Visser's contribution lies in his different interpretations of the pressure and the volume compared to earlier works. The RPST formalism is a version of the Visser thermodynamics in which the cosmological constant is fixed, while the Newton constant is  variable. The RSPT formalism excludes the pressure-volume $(P,V)$ pair and includes $(C,\mu)$ one. In this context, the Smarr relation is a first-order homogeneous for the extensive variables, with mass interpreted as internal energy. This formalism has been subject of many interesting studies, including black holes in higher dimensions  \cite{RP1},  Taub–NUT AdS black holes  \cite{RP2}, black holes in a weak gravity conjecture \cite{RP4},  Einstein-Power-Yang–Mills AdS black holes \cite{RP5} and 4D-EGB black holes \cite{ladghami}.  While the RPST formalism addresses the issues raised in EPST formalism, it does not have any connection to the original black hole chemistry. This is because it assumes the AdS radius to be constant, implying that the size of  CFT remains constant. Consequently, the variation of the central charge  stems from the variation of Newton's constant originating from the bulk \cite{ahm}.
\\

 Based  on the  conformal symmetry of the dual conformal field theory, the authors of \cite{ahm}  introduced a new holographic approach. A pivotal aspect of this approach involves a reevaluation of the Anti-de Sitter boundary conformal factor, $\omega$, considering it as a thermodynamic variable and does not depend on the AdS radius, $L$. This consideration contrasts with the approach taken in Visser's thermodynamics and RPST \cite{vt,gd}, where they employed $\omega = 1$. In another study \cite{H174}, a different choice was made by utilizing a factor $R/L$ instead of the conformal factor $\omega$, with $R$ is the  curvature radius   at the boundary. Further details on these distinctions will be explored in Section \ref{PS1}. Through this setup, they established a new dictionary linking  quantities of AdS black holes with those of the dual conformal field theory. Additionally, they successfully formulated a holographic first law that  corresponds to the first law of extended black hole thermodynamics.  Notably, this approach  circumvents challenges encountered in the EPST formalism, all the while avoiding the need to treat Newton’s constant as a variable. In this research, we call this formalism “conformal holographic extended thermodynamic (CHET)". This new formalism represents an extension of black hole chemistry. More importantly, it offers a comprehensive holographic perspective that can enhance our understanding of black hole phase transitions and their stability. It may also provide insights into the relationship between gravity and quantum physics. Furthermore, it holds the potential for applications in various fields, including cosmology.
\\

In this work, we examine the CHET formalism applied to charged black holes in Anti-de Sitter spacetime  to gain insights into their thermodynamic behavior and phase transitions. We have chosen these black holes  due to their simplicity and significance in the context of black hole thermodynamics. This work, at our knowledge, is the first application of the CHET formalism to black holes.
\\
  
This paper is organized as follows: In Section \ref{PS1}, we present the new dictionary and the construction of conformal holographic extended thermodynamics. In Section \ref{PS2}, we define thermodynamic quantities of charged AdS black holes in the CHET formalism. In Section \ref{PS3}, we study the behavior of thermodynamics and the critical phenomena of charged AdS black holes in CHET. We discuss the results and conclude in Section \ref{PS4}.

In this paper, we adopt the units $\hbar = c = k_B = 1$.
\section{Conformal Holographic Extended Thermodynamics  }
\label{PS1}
In this section, we  review the construction of conformal holographic extended thermodynamics, including the central charge, the new expression for the CFT volume, as well as the first law and Smarr relation. This construction is carried out within the framework of AdS/CFT correspondence.  The cornerstone of this approach is to consider the conformal factor of the Anti-de Sitter boundary as  thermodynamic variable. 
\subsection{Conformal Factor}
In the majority of instances, as in the standard case, the AdS radius, $L$, is typically selected to match the boundary curvature radius of the bulk AdS spacetime, $R$. Under this condition, the  CFT metric can be described as follows \cite{vt}
\begin{equation}
	   \label{C1}
	ds^2= - dt^2 + L^2d \Omega_{d-2}^2,
\end{equation}
where $d\Omega_{d-2}^2$ and $d$  represent   the metric on the round unit $(d - 2)$-sphere and  the dimension number of the bulk, respectively. Additionally, the CFT volume,  $\mathcal{V}$, and the central charge,  $C$, are related as follows
\begin{equation}
	\mathcal{V} \propto L^{d-2} \quad \text{and} \quad C \propto \frac{L^{d-2}}{G},
\end{equation}
where  $G$ is the Newton constant. The definition of the boundary curvature radius, $R=L$, as adopted in the Visser thermodynamics and RPST formalism  \cite{gd,vt}, introduces a key issue. This issue stems from the fact that the central charge and the volume are directly proportional to the AdS radius. Consequently, any alterations made to the radius $L$ will result in corresponding changes to both the central charge $C$ and volume $\mathcal{V}$, rendering them interdependent.  An alternative, broader definition for the  curvature radius at the boundary, i.e. at $R \ne L$, were employed in \cite{H174}, in which the conformal field theory’s metric undergoes a re-scaling, as depicted below
\begin{equation}
	ds^2= \frac{R^2}{L^2}\left( - dt^2 + L^2d \Omega_{d-2}\right).
\end{equation}
This formulation is constructed in order to distinguish the central charge and the volume of the CFT. Hence,  the volume of CFT is proportional to $R^{d-2}$ rather than to the AdS radius. This means that the variation of the central charge and the CFT volume are independent. Belhadj et al. \cite{ahm}, replaced the $R/L$ factor with an arbitrary dimensionless factor, $\omega$, independent of the AdS radius
\begin{equation}
	\label{C3}
	d s^2=\omega^2\left(-d t^2+L^2 d \Omega_{d-2}^2\right).
\end{equation}
This factor, called the conformal factor, reflects the conformal symmetry of  CFT and is regarded as a thermodynamic variable.  In this approach, the volume is redefined by 
\begin{equation}
	\label{1}
	\mathcal{V} \propto(\omega L)^{d-2}.
\end{equation}
This modification exemplifies its innovative approach to addressing the problems aforementioned above.
\subsection{AdS/CFT dictionary}
Based upon the idea of \cite{ahm}, the CFT time used in Eq. \eqref{C1}   differs from the one in the re-scaling case, Eq. \eqref{C3}, by the conformal factor. Indeed, expressing the CFT time by $\tilde{t}$  one has
\begin{equation}
	\tilde{t} = \omega t.
\end{equation}
The CFT re-scaled energy, $\tilde{E}$, is given by the relation
\begin{equation}
	\label{1aa}
\tilde{E} = E \frac{dt}{d\tilde{t}} = \frac{M}{\omega},
\end{equation}
where $M$ denotes the black hole’s energy. We can derive the rest of the CFT quantities from Eq. \eqref{1aa} as
\begin{equation}
	\label{2aa}
		\tilde{T}= \dfrac{\partial \tilde{E}}{\partial S}, \quad \tilde{\Phi}= \dfrac{\partial \tilde{E}}{\partial \tilde{Q}}\quad \text{and} \quad \tilde{\Omega}= \dfrac{\partial \tilde{E}}{\partial J}.
\end{equation}
Here, $\tilde{Q} = Q L/\sqrt{G}$ is the re-scaled electric charge. The quantities of AdS black holes are found through the following equations
\begin{equation}
	\label{3aa}
		T= \dfrac{\partial M}{\partial S}, \quad \Phi= \dfrac{\partial M}{\partial Q}\quad \text{and} \quad \Omega= \dfrac{\partial M}{\partial J}.
\end{equation}
From Eqs. \eqref{1aa}-\eqref{3aa},  the  holographic dictionary relating the quantities of AdS black holes (without tildes) to the quantities of CFT (with tildes) is as follows \cite{ahm}
\begin{equation}
	\label{3}
	\begin{aligned}
		&\tilde{S}=S=\frac{A}{4 G}, \quad \tilde{E}=\frac{M}{\omega}, \quad \tilde{T}=\frac{T}{\omega}, \
		&\tilde{\Omega}=\frac{\Omega}{\omega}, \quad \tilde{J}=J, \quad \tilde{\Phi}=\frac{\Phi \sqrt{G}}{\omega L}, \quad \tilde{Q}=\frac{Q L}{\sqrt{G}} .
	\end{aligned}
\end{equation}
where $S$,  $T$, $M$, $\Omega$, $J$, $\Phi$, $Q$, $\tilde{E}$ are the Bekenstein-Hawking
entropy, Hawking temperature, black hole mass, the angular velocity, the angular velocity, the electrical potential, the electric charge, the CFT energy, respectively.  In this approach, the new dictionary is the most general, and the mass of black holes is proportional to internal energy by the conformal factor, in contrast to the old interpretation in EPST , where it is interpreted as enthalpy \cite{Kastor,7n}, or in the RPST formalism, where the mass of black holes equals the internal energy \cite{gd}. However, the entropy is invariant under this re-scaling, which means that the re-scaling does not affect the partition function of the boundary theory. When the CFT metric is re-scaling, it leads to a re-scaling of CFT time, $t\to \omega t$, and of the  most CFT quantities, including internal energy and temperature. 	

\subsection{First Law and Smarr Relation }
We can get the first law of conformal holography extended thermodynamics, by re-scaling with the conformal factor and treating it as a thermodynamic variable, where the derivation of internal energy is expressed by
\begin{equation}
	\label{7ppp}
	d\left( \frac{M}{\omega}\right) = \frac{d M}{\omega} - \frac{M}{\omega^2}\, d\, \omega
\end{equation}
where $d M$ is given by \cite{vt}
\begin{equation}
	\label{8ppp}
	\begin{aligned}
		d M= & T d\left(\frac{A}{4 G}\right)+\Omega d J+\frac{\Phi}{L} d(Q L)-\frac{M}{d-2} \frac{d L^{d-2}}{L^{d-2}} \\
		& +\left(M-T S-\Omega J-\frac{\Phi}{L} Q L\right) \frac{d\left(L^{d-2} / G\right)}{L^{d-2} / G}.
	\end{aligned}
\end{equation}
By using Eqs. \eqref{7ppp} and \eqref{8ppp}, we can derive the first law as follows \cite{ahm}
\begin{equation}
	\label{4}
	\begin{aligned}
	d \left(\frac{M}{\omega}\right)= & \frac{T}{\omega} d \left(\frac{A}{4 G}\right)+\frac{\Omega}{\omega} d J+\frac{\Phi \sqrt{G}}{\omega L} d \left(\frac{Q L}{\sqrt{G}}\right) \\
		& +\left(\frac{M}{\omega}-\frac{T S}{\omega}-\frac{\Omega J}{\omega}-\frac{\Phi Q}{\omega}\right) \frac{d \left(L^{d-2} / G\right)}{L^{d-2} / G}  -\frac{M}{\omega(d-2)} \frac{d (\omega L)^{d-2}}{(\omega L)^{d-2}}.
	\end{aligned}
\end{equation}
Using Eqs. \eqref{1}-\eqref{3},  we simplify the holography first law as
\begin{equation}
	\label{first}
	d \tilde{E}=\tilde{T} d S+\tilde{\Omega}d J+\tilde{\Phi} d \tilde{Q}+\mu d C-p d \mathcal{V}
\end{equation}
where the chemical potential $\mu$ is    
\begin{equation}
	\label{74}
	\mu=\frac{1}{C}(\tilde{E}-\tilde{T} S-\tilde{\Omega} J-\tilde{\Phi} \tilde{Q})
\end{equation}
and
\begin{equation}
	\label{00}
 (d-2)	p \mathcal{V}=\tilde{E}.
\end{equation}
The last equation represents an equation of state (EoS). To eliminate the  term $-p d\mathcal{V}$ from the first law, Eq. \eqref{first}, we re-scale certain CFT quantities by a factor of $\omega L$ using the following formulas \cite{ahm}
\begin{equation}
	\label{04}
	\hat{E}=\omega L \tilde{E}, \quad \hat{T}=\omega L \tilde{T}, \quad \hat{\Omega}=\omega L \tilde{\Omega}, \quad
\hat{\Phi}=\omega L \tilde{\Phi}, \quad \hat{\mu}=\omega L \mu .
\end{equation}
We obtain a first law that does not contain pressure and volume \cite{ahm}
\begin{equation}
	\label{y0}
d \hat{E}=\hat{T} d S+\hat{\Omega} d J+\hat{\Phi} d \tilde{Q}+\hat{\mu} d C.
\end{equation}
By using Eqs. \eqref{74} and \eqref{04}, we derive the Smarr relation \cite{ahm}
\begin{equation}
	\label{y1}
	\hat{E}=\hat{T} S+\hat{\Omega} J+\hat{\Phi} \tilde{Q}+\hat{\mu} C.
\end{equation}
Conformal holographic extended thermodynamics is based on three key ideas. The first idea involves a re-scaling of the CFT metric using the conformal factor. The second idea is to treat the AdS radius as a thermodynamic variable while keeping Newton constant fixed. The final idea is to regard the conformal factor as a thermodynamic parameter independent of the AdS radius. This framework yields a fresh formulation for the CFT volume and prevents the degeneration of the first law, allowing the central charge and the CFT volume to vary independently, without adopting the approach proposed in the RPST formalism. In Smarr relation, the internal energy is a first-order homogeneous function as RPST formalism.
\\

\section{Charged AdS Black hole}
\label{PS2}
In this section, we use the results of the conformal holographic extended thermodynamic to define thermodynamic quantities  of a Reissner-Nordström black holes in AdS spacetime (RN-AdS black holes). The metric of this  black hole is 
\begin{equation}
 \mathrm{d} s^2=-f(r) \mathrm{d} t^2+f(r)^{-1} \mathrm{d} r^2+r^2\left(\mathrm{d}  \theta ^2+\sin ^2 \theta \mathrm{d} \phi^2\right),
\end{equation}
where the metric function $f(r)$ is 
\begin{equation}
	\label{metric}
	f(r)=1-\frac{2 M  G}{r}+\frac{ Q^2  G}{r^2}+\frac{r^2}{L^2},
\end{equation}
where M and Q are the mass and the electrical charge of black holes, respectively. 
By solving the equation $f(r) = 0$, two solutions are found, $r_{\pm}$, where $r_+$  corresponds to the event horizon of the black hole and $r_- $ is the Cauchy horizon. For $r = r_+$, the expression of black holes mass write
\begin{equation}
	\label{Mass}
	M=\frac{r_{+}}{2 G}\left(1+\frac{G Q^2}{r_{+}^2}+\frac{r_{+}^2}{L^2}\right).
\end{equation}
To construct holographic extended thermodynamics of RN-AdS black holes, we need to determine the thermodynamic quantities of this black hole. To this aim, the RN-AdS black holes entropy, the electrical potential and the Hawking temperature at the  event horizon  are given by 
\begin{equation}
	\label{Entropy}
S=\frac{A}{4 G}=\frac{\pi r_{+}^2}{G},
\end{equation}

\begin{equation}
	\Phi = \frac{Q}{r_+},
\end{equation} 
and
\begin{equation}
	T=\frac{f^{\prime}\left(r_{+}\right)}{4 \pi}=\frac{r_+^2+ 3 r_+^4 L^{-2}-G Q^2}{4 \pi r_+^3},
\end{equation}
respectively. By using the new dictionary between  quantities of AdS black holes and CFT, Eqs.\eqref{3} and \eqref{04}, we find the CFT quantities that correspond to those of the  RN-AdS black hole as follow
\begin{equation}
		\label{074}
	\hat{E} \equiv\hat{M}=  M\, L, \quad \hat{T}=  T\,L, \quad \tilde{S}=S,\quad \tilde{Q}=\frac{Q L}{\sqrt{G}}, \quad
	\hat{\Phi}=\Phi \sqrt{G},\quad C=\frac{L^2}{G} ,\quad \hat{\mu}=\omega L \mu .
\end{equation}
From Eqs. \eqref{y0} and \eqref{y1}, the first law  and Smarr relation in CHET of the charged AdS black holes write as
\begin{equation}
	\label{y7}
	d \hat{M}=\hat{T} d S+\hat{\Phi} d \tilde{Q}+\hat{\mu} d C,
\end{equation}
and
\begin{equation}
	\label{y8}
	\hat{M}=\hat{T} S+\hat{\Phi} \tilde{Q}+\hat{\mu} C,
\end{equation}
respectively. Finally, we get thermodynamic variables from the first law, Eq. \eqref{y8}, and the mass of black holes, Eq. \eqref{Mass}, as
\begin{equation}
	\label{T}
	\hat{T}=\left(\frac{\partial \hat{M}}{\partial S}\right)_{\tilde{Q}, C}= \frac{\pi  C S-\pi ^2\tilde{Q}^2+3 S^2}{4 \pi ^{3/2} S \sqrt{C S}},
\end{equation}
\begin{equation}
\hat{\Phi}=\left(\frac{\partial \hat{M}}{\partial \tilde{Q}}\right)_{S, C}=	\frac{\sqrt{\pi } \tilde{Q}}{\sqrt{C S}},
\end{equation}
\begin{equation}
	\label{Po}
\hat{\mu}=\left(\frac{\partial \hat{M}}{\partial C}\right)_{S, \tilde{Q}}=	\frac{\pi  C S-\pi ^2 \tilde{Q}^2-S^2}{4 \pi ^{3/2} C \sqrt{C S}}.
\end{equation}
\section{Phase structure}
\label{PS3}
\subsection{Critical phenomena}
In this section, we study the phase transition and the critical central charge  for the charged AdS black holes in the conformal extended holographic thermodynamic. To find the coordinate of  the critical point i.e. the event horizon and the critical central charge, we solve the following system
\begin{equation}
	\left(\frac{\partial \hat{T}}{\partial S}\right)_{\tilde{Q}, C}=0 \quad\text{and} \quad \left(\frac{\partial^2 \hat{T}}{\partial S^2}\right)_{\tilde{Q}, C} =0,
\end{equation}
one gets
\begin{equation}
	S_c= \pi\, \tilde{Q}, \qquad C_c=6\, \tilde{Q},\qquad \tilde{T}_c= \frac{1}{\pi} \sqrt{\frac{2}{3}}.
\end{equation}
Using the following normalization
\begin{equation}
	\label{No}
	t= \frac{\tilde{T}}{\tilde{T}_c}, \qquad s = \frac{S}{S_c},  \qquad c = \frac{C}{C_c},
\end{equation}
we can rewrite the equation of state, Eq. \eqref{T}, in the following form
\begin{equation}
	t= \frac{3s^2 + 6s\,c -1}{8s\,\sqrt{s\,c}}.
\end{equation}
\begin{figure}[htp]
	\label{1p}
	\centering
	\includegraphics[scale=0.7]{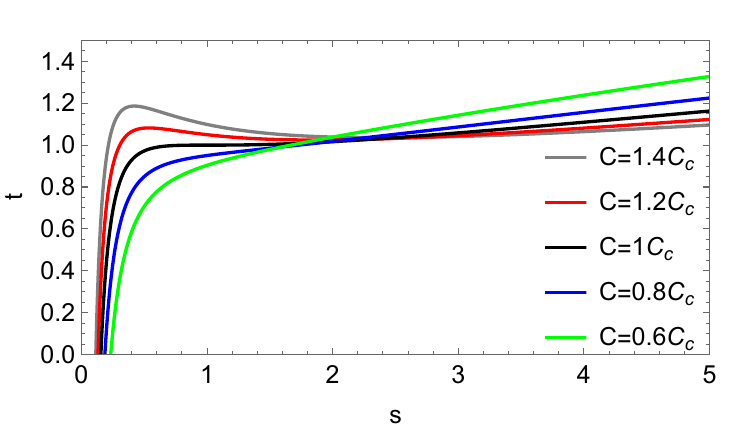}
	\caption{t - s  curves in the iso-central-charge processes}
	\label{Tux}
\end{figure}
\\

Fig. \ref{Tux} shows  the evolution of the Hawking temperature in terms of the  entropy or the event horizon of the black hole, for different values of the central charge. For $C=C_c$, a phase transition of the second order  occurs between a small and a large black hole. For $C>C_c$, we notice a non-monotonous curves,  where  a first order phase transition occurs, between a large and a small black hole, passing through a medium black hole. Finally, no phase transition  occurs for $C<C_c$. This thermodynamic behavior is similar to that of the Van der Waals fluids, as well as to that of the thermodynamic behavior for RN-AdS black holes in the extended phase space thermodynamics \cite{EPST} and the restricted phase space thermodynamics \cite{gd}. The main discrepancy resides in the effect of the central charge on the thermodynamic behavior rather than the effect of the electrical charge. 
\subsection{Free Energy}
	In the context of the phase structure of RN-AdS black holes in the CHET formalism, we can also study the thermodynamic behavior in terms of the evolution of the free energy and its relation to the dual CFT. We express the Helmholtz free energy $\hat{F}$ in conformal holography extended thermodynamics by
\begin{equation}
	\label{..}
	\hat{F}= \hat{M} - \hat{T}S
\end{equation}
 By normalizing the free energy as $f= \hat{F}/ \hat{F}_c$, where $\hat{F}_c= \sqrt{\frac{2}{3}} Q$,  Eq. \eqref{..} becomes
\begin{equation}
	f=\frac{s \left(-4 \sqrt{c s}\, t+6 c+s\right)+1}{4 \sqrt{c s}}.
\end{equation}
\begin{figure}[htp]
	\centering
	\includegraphics[scale=0.6]{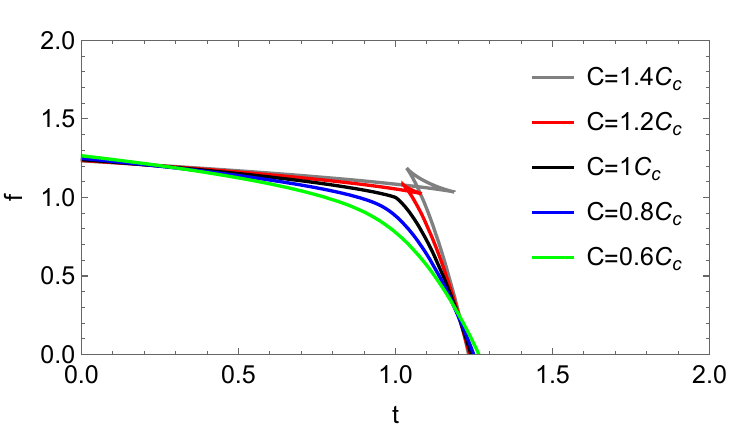}
	\caption{f - t curves in the iso-central-charge processes}
	\label{libre}
\end{figure}
 Fig. \ref{libre} represents the evolution of the normalized Helmholtz free energy in terms of the normalized  temperature for different values of the central charge. In the free energy evolution curve, a swallow's tail appears for a central charge greater than the critical one corresponding  to a first-order phase transition. This swallow's tail disappears for a central charge  equals or less than the critical central charge. 
\subsection{Heat Capacity} 
The heat capacity gives us an idea of the stability/instability of black holes. It is known that the system is stable (unstable) for a positive (negative) heat capacity. The heat capacity is expressed by 
\begin{equation}
	C_{\tilde{Q}} = \hat{T} \left( \frac{\partial{S}}{\partial{\hat{T}}}\right)_{\tilde{Q}} = \frac{2 S \left(\pi  C S-\pi ^2 \tilde{Q}^2+3 S^2\right)}{3 S^2-\pi  C S+3 \pi ^2 \tilde{Q}^2}.
\end{equation}
We normalize the heat capacity by
\begin{equation}
	\eta = \frac{C_{\tilde{Q}}}{2 \pi \tilde{Q}} = \frac{3s\left(s+2sc \right) -s}{3 s (s-2 c)+3}.
\end{equation}
\begin{figure}[htp]
	\centering
	\includegraphics[scale=0.40]{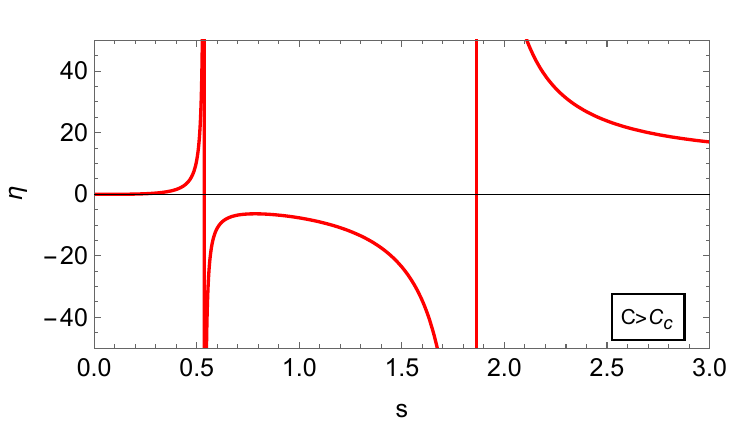}
	\includegraphics[scale=0.40]{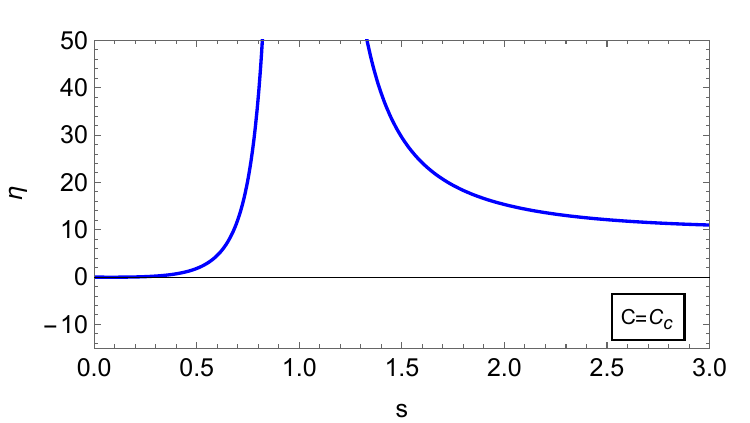}
	\includegraphics[scale=0.40]{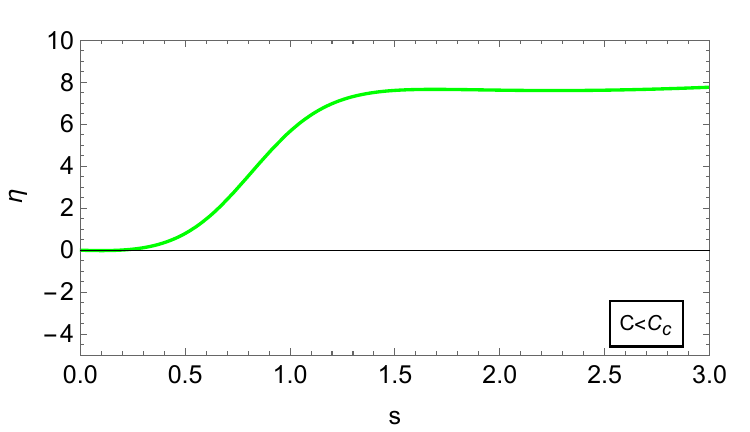}
	\caption{Heat capacity in term of the entropy for different values of C.}
	\label{2pp}
\end{figure}
Fig. \ref{2pp} represents the normalized heat capacity in terms of the normalized entropy for different values of the central charge. From Figs. \ref{Tux} and \ref{2pp}, we conclude that small and large black holes are stable, while  medium black holes are unstable. The thermodynamic behavior of black holes depends on the dual CFT i.e. on the central charge. Indeed, for a large central charge ($C>6\hat{Q}$),  a phase transition between small and large black holes is observed. While for a small central charge no phase transition occurs.
\subsection{Chemical Potential}
In statistical physics of ordinary matter, the sign of chemical potential plays a role in determining the type of regime we are studying, i.e. classical or quantum regime. By analogy between statistical physics of ordinary matter and conformal holography extended thermodynamics of AdS black holes, we define  conditions on the central charge C to distinguish  quantum regime from classical one.
 If the chemical potential is positive, then the regime corresponds to a quantum regime and if it is negative, then it corresponds to a classical regime \cite{P}. Using  Eqs. \eqref{04}, \eqref{Po} and \eqref{No}, we find 
\begin{equation}
	\hat{\mu} = \frac{6\, c\, s - s^{2} - 1}{24 \,c\,\sqrt{6\,c\,s}},
\end{equation}
 Thus, the system corresponds to a quantum regime for
\begin{equation}
	C > \tilde{Q} \left(s + \frac{1}{s} \right),
\end{equation}
and  to a classical one for the case  of  $	C < \hat{Q} \left(s + s^{-1} \right)$. The chemical potential is zero for $C = \hat{Q} \left(s + s^{-1} \right)$.  This condition may be interpreted as a non holographic point as   the element $\hat{\mu}dC$ and $\hat{\mu}C$  disappear from  the first law and Smarr relation, respectively.  In other words, we recover the traditional  black hole thermodynamic (TBHT) formalism.
\begin{figure}[htp]
	\centering
	\includegraphics[scale=0.70]{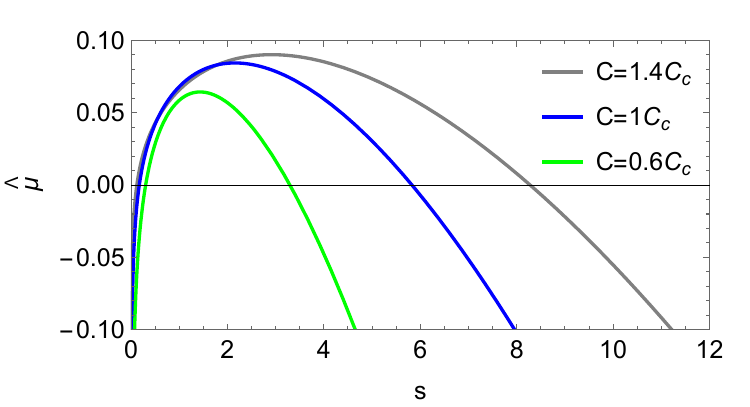}
		\caption{Chemical potential in term of entropy for different values of C.}
	\label{3pp}
\end{figure}
\\

By comparing Figs. \ref{Tux}, \ref{2pp} and \ref{3pp}, we conclude that the phases transition region are located in the quantum regime i.e. for $\hat{\mu}>0$ and $0.5<s<2$, while for the classical regime, i.e.  $\hat{\mu}<0$ and large entropy, there is no phases transition.
For the Schwarzschild-AdS black hole ($\tilde{Q}=0$) we find that the quantum regime , from Eq. \eqref{Po},  is characterized by
  \begin{equation}
  	S< \pi C \quad 
  \end{equation}
 and the classical regime by  
   \begin{equation}
  	S> \pi C \quad 
  \end{equation}

 \section{Discussion and Conclusion }
 \label{PS4}
 \paragraph{}
 In this work, we have studied a charged AdS black holes in the conformal holography extended thermodynamic by using the new dictionary between  quantities of the extended black hole thermodynamic and the quantities of the dual CFT in the context of the AdS/CFT correspondence. Through this dictionary, we get the first law and Smarr relation of the conformal holography extended thermodynamic formalism. In this formalism, the interpretation of black hole mass differs from the previous formalisms i.e. TBHT, EPST, and RPST.  In CHET, the mass is a homogeneous function of the first-order without resorting to making Newton constant $G$ a variable as it will be in the RPST formalism. By re-scaling some quantities of  CFT,  the pressure-volume disappear from the first law. Furthermore, in this formalism, the volume is not the conjugate quantity of the thermodynamic pressure of the black hole as in the EPST formalism, but is the CFT volume from which a new pressure is defined as its conjugate quantity.
 \paragraph{}
   By using the results of the CHET formalism, we have  studied the  thermodynamic  behavior and the  property  of a charged AdS black hole  and  its relationship to the states of the dual CFT for a fixed electrical charge. We found expressions of the  thermodynamic quantities  and the equation of state in this approach. We have also studied the effect of the CFT's states for the thermal evolution of black holes. For a critical central charge, $C=C_c$, an emergency second-order phase transition appears between a  small and a large black hole. For the large central charge, i.e. $C>C_c$, we find a first-order phase transition between  small and  large black holes, passing through an unstable medium black hole with a negative heat capacity, in contrast to large and small black holes, which are stable system, as shown in Figs. \ref{Tux} and \ref{2pp}. Finally, no phase transitions occurs for a small central charge ($C<C_c$). This thermodynamic behavior  is similar to that of Van der Waals fluids. 
  After normalizing the equation of state, the electric charge no longer exists. Indeed, the value of the electric charge has no effect on the thermodynamic behavior of charged AdS black holes, but this does not mean that the critical phenomena are achieved in an electrically neutral black hole. Whereas, the state of CFT depends on the electric charge as $C_c=6\tilde{Q}$ and discussed above.
  \paragraph{}
  Finally, we have studied the regime of the system in terms of the chemical potential. We find that the case of $	C > \hat{Q} \left(s + s^{-1} \right)$ corresponds to a quantum regime while the case of $	C < \hat{Q} \left(s + s^{-1} \right)$ corresponds to the classical one. We notice that the non holographic point,  $	C = \hat{Q} \left(s + s^{-1} \right)$, corresponds to the TBHT formalism.


\end{document}